\documentclass[conference]{IEEEtran}
\IEEEoverridecommandlockouts
\usepackage{balance}
\usepackage{cite}
\usepackage{amsmath,amssymb,amsfonts}
\usepackage[top=0.71in,bottom=1in,right=0.625in,left=0.625in]{geometry}
\usepackage{array}
\usepackage[caption=false,font=footnotesize,position=top]{subfig}
\usepackage{stfloats}
\usepackage{url}
\usepackage{graphicx}
\usepackage{textcomp}
\usepackage{xcolor}
\usepackage{tikz}
\usepackage{ellipsis}
\usetikzlibrary{calc}
\usetikzlibrary{decorations.pathreplacing,decorations.markings,shapes.geometric}
\usetikzlibrary{calc,patterns,angles,quotes,shapes,arrows.meta}
\usetikzlibrary{chains}
\usepackage{multirow}

\usepackage[utf8]{inputenc}

\usepackage{mathtools}
\usepackage{bm}
\usepackage{etoolbox}
\usepackage{scalerel}

\usepackage{fancyhdr}

\usepackage{makecell}

\newcommand{\va}{{\bm{a}}}

\newcommand{\vh}{{\bm{h}}}

\newcommand{\vn}{{\bm{n}}}

\newcommand{\vs}{{\bm{s}}}

\newcommand{\vy}{{\bm{y}}}

\newcommand{\ma}{{\bm{A}}}

\newcommand{\mc}{{\bm{C}}}

\newcommand{\matp}{{\bm{P}}}

\newcommand{\mt}{{\bm{T}}}

\newcommand{\mv}{{\bm{V}}}

\newcommand{\my}{{\bm{Y}}}

\newcommand{\vth}{\bm{\theta}}

\newcommand{\vmu}{\bm{\mu}}
\newcommand{\vdel}{{\bm{\delta}}}

\newcommand{\He}{\mathrm{H}}
\newcommand{\T}{\mathrm{T}}

\newcommand{\I}{\mathbf{I}}
\renewcommand{\j}{\mathrm{j}}
\newcommand{\NC}{\mathcal{N}_\mathbb{C}}

\tikzstyle{block} = [draw, rectangle, 
minimum height=4em, minimum width=4em]
\tikzstyle{input} = [coordinate]
\tikzstyle{output} = [coordinate]
\tikzstyle{pinstyle} = [pin edge={to-,thin,black}]

\usetikzlibrary{positioning}

\tikzset{radiation/.style={{decorate,decoration={expanding waves,angle=90,segment length=5pt}}}}

\usetikzlibrary{spy}
\usepackage{pgfplots}
\usepgfplotslibrary{fillbetween}
\usepackage{wrapfig}
\usetikzlibrary{arrows,shapes}
\usetikzlibrary{positioning,shapes.callouts}
\usepgfplotslibrary{groupplots,dateplot}
\usetikzlibrary{patterns,shapes.arrows}
\pgfplotsset{compat=newest}

\def\BibTeX{{\rm B\kern-.05em{\sc i\kern-.025em b}\kern-.08em
		T\kern-.1667em\lower.7ex\hbox{E}\kern-.125emX}}

\usepackage[acronym,shortcuts]{glossaries}
\newacronym{GMM}{GMM}{Gaussian mixture model}
\newacronym{PDF}{PDF}{probability density function}
\newacronym{MSE}{MSE}{mean square error}
\newacronym{CSI}{CSI}{channel state information}
\newacronym{CME}{CME}{conditional mean estimator}
\newacronym{ML}{ML}{maximum likelihood}
\newacronym{LS}{LS}{least squares}
\newacronym{LOS}{LoS}{line-of-sight}
\newacronym{NLOS}{NLoS}{non-\ac{LOS}}
\newacronym{DoA}{DoA}{direction-of-arrival}
\newacronym{SNR}{SNR}{signal-to-noise ratio}
\newacronym{BS}{BS}{base station}
\newacronym{JCAS}{JCAS}{joint communication and sensing}
\newacronym{MIMO}{MIMO}{multiple-input-multiple-output}
\newacronym{MMSE}{MMSE}{minimum \ac{MSE}}
\newacronym{NMSE}{NMSE}{normalized \ac{MSE}}
\newacronym{RMSE}{RMSE}{root \ac{MSE}}
\newacronym{LMMSE}{LMMSE}{linear \ac{MMSE}}
\newacronym{MT}{MT}{mobile terminal}
\newacronym{UE}{UE}{user equipment}
\newacronym{OMP}{OMP}{orthogonal matching pursuit}
\newacronym{CS}{CS}{compressed sensing}
\newacronym{ULA}{ULA}{uniform linear array}
\newacronym{DFT}{DFT}{Discrete Fourier Transform}
\newacronym{MUSIC}{MUSIC}{multiple signal classification}
\newacronym{ESPRIT}{ESPRIT}{estimation of signal parameters via rotational invariance techniques}
\newacronym{GE}{GE}{gridded estimator}
\newacronym{AWGN}{AWGN}{additive white Gaussian noise}
\newacronym{GSC}{GSC}{generalized sidelobe cancellor}
\newacronym{wlog}{w.l.o.g.}{without loss of generality}

\fancypagestyle{cfooter}{ %
	\fancyhf{} 
	\cfoot{
		This work has been submitted to the IEEE for possible publication. Copyright may be transferred without notice, after which this version may no longer be accessible.}
	
}

\begin{document}

\title{
DoA-Aided MMSE Channel Estimation \\for Wireless Communication Systems
\thanks{This work was supported by the Federal Ministry of Education and Research of Germany in the programme of “Souverän. Digital. Vernetzt.”. Joint project 6G-life, project identification number: 16KISK002}
}
\newcommand{\jbig}{$\mathcal{J}$}
\author{\IEEEauthorblockN{Franz Weißer, Nurettin Turan, and Wolfgang Utschick\\}
\IEEEauthorblockA{\textit{TUM School of Computation, Information and Technology, Technical University of Munich, Germany} \\
\{franz.weisser, nurettin.turan, utschick\}@tum.de}
}

\maketitle
\thispagestyle{cfooter}

\begin{abstract}
	This paper investigates the combination of parametric channel estimation with minimum mean square error (MMSE) estimation.
We propose a direction-of-arrival (DoA)-aided two-stage channel estimation technique that utilizes the decomposition of wireless communication channels into a line-of-sight (LoS) path and its orthogonal subspace.
After estimating the channel along the dominant direction,
we utilize a Gaussian mixture model to estimate the conditionally Gaussian distributed random vector, which represents the multipath propagation.
The proposed two-stage estimator allows pre-computing the respective estimation filters, tremendously reducing the computational complexity.
Numerical simulations with typical channel models depict the superior performance of our proposed two-stage estimation approach compared to state-of-the-art methods.
\end{abstract}


\section{Introduction}

With the need for higher data rates and more reliable transmissions, channel estimation has become a crucial task in modern \ac{MIMO} communication systems~\cite{Rusek2013}.
Similar to other fields of wireless communication, machine learning-based methods have been introduced for channel estimation~\cite{Soltani2019,Neumann2018,Koller2022}.
Based on a representative training data set, machine learning-based methods learn a prior distribution during the learning stage, which enhances the channel estimation performance later on.
Notably, the works of~\cite{Neumann2018,Koller2022} propose channel estimation methods that approximate the \ac{MMSE} estimator, focusing on the statistics of the channels without utilizing any structural information given by the geometric channel model.

As an alternative approach, parametric channel estimators have been proposed, which build on the multipath propagation characteristics of wireless channels, e.g.,~\cite{Shafin2016,Larsen2009}. 
These so-called model-based techniques have gained relevance in recent years, estimating the parameters of each path and constructing the channel based on these estimates.
Those parametric channel estimators are especially relevant when considering \ac{JCAS} systems.
\ac{JCAS} systems aim to integrate communication and sensing functionalities in one system to enhance overall capabilities.
In~\cite{Liu2022}, a comprehensive overview of the progress in \ac{JCAS} research is given.

In~\cite{Shafin2016}, it is shown that \ac{DoA}-based channel estimation can achieve superior performance compared to the \ac{LMMSE} estimation, while the authors in~\cite{Larsen2009} evaluate performance bounds, validating the superiority of parametric channel estimation.
Notably, these works focus primarily on ray-based channel models, describing each multipath component by a distinct ray stemming from a point scatterer.
These models are often assumed for high frequencies. 
However, for sub $6$ GHz, realistic channel models, e.g.,~\cite{3GPP2020}, model scatterers not as points but rather as point clouds resulting in propagation clusters composed of a dominant path and accompanying subpaths, which may be not resolved by distinct rays.
Studies on the impact of subpaths for higher frequencies, such as mmWave are ongoing research, cf.~\cite{Antonescu2019}.
The estimation of these subpaths is crucial for the communication task since they allow the full utilization of the achievable data rate~\cite{Tse2005}.
In the context of massive \ac{MIMO} systems, the substantially improved spatial resolution offered by large-scale antenna arrays has led to the employment of spatial channel models \cite{Yin2013,Neumann2018}.

\emph{Contributions:} 
This work aims to bridge the gap between parametric and \ac{MMSE}-based channel estimation.
To this end, we reformulate the estimation problem similar to the \ac{GSC} into \ac{LOS} and \ac{NLOS} subspaces based on \ac{DoA} estimates either provided by localization and positioning functionalities or established \ac{DoA} estimators.
Afterwards, we introduce two independent estimators, where the estimation of the \ac{NLOS} components utilizes the channel estimator based on a \ac{GMM}, as proposed in~\cite{Koller2022}. 
This combination seeks to enhance the accuracy and efficiency of channel estimation in realistic wireless communication scenarios.

\section{System Model}
\label{sec:system_model}

\begin{figure}
	\centering
	\begin{tikzpicture}[scale=1, decoration={markings, mark= at position 0.75 with {\arrow{stealth}}}] 
		\draw[fill] (5,1.3) circle (2pt) coordinate (Tx2);
		
		\node[right] at (Tx2) {\footnotesize UE};
		
		\draw[fill] (0,0) circle (1pt) coordinate (Rx);
		
		\draw[fill] (Rx)+(0.45, -0.45) circle (1pt);
		\draw[fill] (Rx)+(0.15, -0.15) circle (1pt);
		\draw[fill] (Rx)+(-0.15, 0.15) circle (1pt);
		\draw[fill] (Rx)+(-0.45, 0.45) circle (1pt);
		\draw[fill] (Rx)+(-0.3, 0.3) circle (1pt);
		\draw[fill] (Rx)+(0.3, -0.3) circle (1pt);
		
		\node[below left] at (Rx) {\footnotesize BS};
		
		\draw[postaction={decorate}] (Tx2) -- (Rx);
		
		\coordinate (a) at (2.8,2.2);
		\draw plot[domain=0:350, smooth cycle,shift=(a)] (\x:0.3+rnd*0.15 and 0.2+rnd*0.1);
		\coordinate (b) at (6,0);
		\draw plot[domain=0:350, smooth cycle,shift=(b)] (\x:0.2+rnd*0.15 and 0.3+rnd*0.15);
		\coordinate (c) at (5.2,2.3);
		\draw plot[domain=0:350, smooth cycle,shift=(c)] (\x:0.2+rnd*0.1 and 0.2+rnd*0.15);
		
		\draw[dashed,name path = A] (Tx2) -- (2.9,2.05) -- (Rx);
		\draw[dashed,name path = B] (Tx2) -- (2.6,2.3) -- (Rx);	
		\draw[postaction={decorate}] (Tx2) -- (a) -- (Rx);	
		
		\draw[dashed,name path = C] (Tx2) -- (6,0.2) -- (Rx);	
		\draw[dashed,name path = D] (Tx2) -- (6,-0.2) -- (Rx);		
		\draw[postaction={decorate}] (Tx2) -- (b) -- (Rx);		
		
		\draw[dashed,name path = E] (Tx2) -- (5.3,2.15) -- (Rx);	
		\draw[dashed,name path = F] (Tx2) -- (5,2.4) -- (Rx);		
		\draw[postaction={decorate}] (Tx2) -- (c) -- (Rx);

	\end{tikzpicture}

\caption{Illustration of a wireless channel between the user equipment (UE) and the base station (BS) with one \ac{LOS} path and three scattering clusters.}
 \vspace{-0mm}
\label{fig:system_model}
\end{figure}
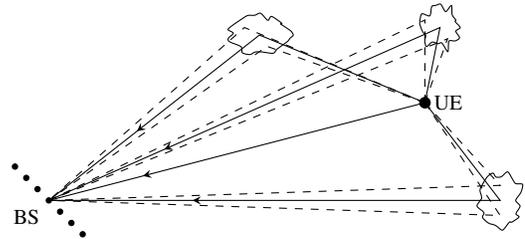

Let us consider a \ac{BS}, equipped with $M$ antennas that receives uplink signals from a single-antenna \ac{UE}.
We assume the \ac{UE} is located in the far field, and we have a frequency-flat, block-fading channel.
Under the assumption that orthogonal pilots are used, we can correlate the received pilot observations with the transmitted signals, such that only the channel of interest is present in the observations.
The received signal vector at the \ac{BS} can be written as
\begin{align}
	\vy(t) = \vh(t) + \vn(t), \quad t=1,...,T, \label{eq:sys_model}
\end{align}
with the channel vectors $\vh(t)$ and the \ac{AWGN} $\vn(t) \sim \NC(\bm{0},\mc_\vn=\sigma^2\I_M)$.
Additionally, the following Rician model is assumed~\cite{3GPP2020,Oezdogan2019}
\begin{align}
	\vh(t) &= \sqrt{\frac{KM}{K+1}} \va(\theta)e^{\j\phi(t)} + \sqrt{\frac{M}{K+1}} \vh_\text{NLoS}(t) \label{eq:chan_model} 
\end{align}
where $K$ denotes the Rician factor, 
$\va(\theta)$ and $\phi(t)$ are the steering vector and phase of the dominant path, and $\vh_\text{NLoS}$ denotes the channel corresponding to the scatterers.
Each channel is normalized by its path loss.
Based on a set of parameters $\vdel$, which describe the directions and properties of the clusters, the \ac{NLOS} channel vectors are considered as conditionally Gaussian distributed~\cite{3GPP2020,Neumann2018}
\begin{align}
	\vh_\text{NLoS}(t) \mid \vdel \sim \NC \left(\bm{0}, \mc_\text{NLoS}(\vdel)\right).
\end{align}
The spatial covariance matrix of the NLoS components is given by
\begin{align}
	\mc_\text{NLoS}(\vdel) = \int_{-\pi}^{\pi} g(\vartheta,\vdel) \va(\vartheta)\va^\He(\vartheta) d\vartheta,
\end{align}
where $g(\vartheta,\vdel)$ is the power density consisting of a weighted sum of Laplace densities, which have standard deviations $\sigma_\text{AS}$ corresponding to the angular spread of the propagation clusters. 
The specifications in~\cite{3GPP2020} for channel simulations include an angular spread of $\sigma_\text{AS}=2^\circ$ and $\sigma_\text{AS}=5^\circ$ for the \ac{NLOS} clusters in an urban macro and urban micro cell, respectively.
For further details on the used spatial channel model, the reader is referred to~\cite{Neumann2018}.
This work considers the channel parameters to be constant within one coherence interval, i.e., $\vdel$ and $\theta$ do not change over the $T$ observations.
The channel covariance matrix of $\vh(t)$ can then be written as
\begin{align}
	\mc_{\vdel,\theta} = \frac{KM}{K+1}\va(\theta)\va^\He(\theta) + \frac{M}{K+1}\mc_\text{NLoS}(\vdel). \label{eq:cov}
\end{align}
If the \ac{BS} employs a \ac{ULA}, the steering vector is given as
\begin{align}
	\va(\theta) = \frac{1}{\sqrt{M}}\left[1, e^{-\j2\pi\frac{d}{\lambda}\sin(\theta)}, ..., e^{-\j2\pi\frac{d}{\lambda}(M-1)\sin(\theta)}\right]^\T, \label{eq:spaceSteering}
\end{align}
where $d$ and $\lambda$ are the antenna element spacing and the carrier frequency wavelength, respectively.

We estimate the most recent channel realization $\vh(T)$, where we will drop the index from now on for notational convenience. 
Depending on the channel estimation method, we either infer the channel solely from the most recent observation $\vy(T)$, or we utilize all observations $\vy(t), t=1,...,T$.
In the former case, we drop the index again for convenience.

\section{Proposed Method}

In order to utilize structural information given as additional prior information about the channel, we propose a \ac{DoA}-aided two-stage approach, making the channel estimate more robust against noise.
Given an observation $\vy$ and the \ac{DoA} of the dominant path denoted by $\theta$ the \ac{MSE} optimal estimator of $\vh$, i.e., \ac{MMSE}-achieving, is given by the \ac{CME}
\begin{align}
	\hat{\vh} = \mathbb{E}[\vh\mid\vy,\theta].
\end{align}

To incorporate this prior information, we make use of a two-stage approach.
Similarly to the \ac{GSC}~\cite{Griffiths1982}, we split the estimation problem into two parts. 
To this end, we rewrite the observation as
\begin{align}
	\Tilde{\vy} = \mt^\He(\theta) \vy = 
	\begin{bmatrix}
		\va^\He(\theta)\\
		\mv^\He(\theta)
	\end{bmatrix}
	\vy = 
	\begin{bmatrix}
		\Tilde{y}_\va \\
		\Tilde{\vy}_{\bar{\va}}
	\end{bmatrix}
	=
	\begin{bmatrix}
		\Tilde{h}_\va \\
		\Tilde{\vh}_{\bar{\va}}
	\end{bmatrix} + \vn
	, \label{eq:transform}
\end{align}
with $\mv^\He(\theta)\va(\theta) = \bm{0}$. 
Consequently, we can denote with ${{\vh}}_\va=\va(\theta)\tilde{h}_\va$ and ${{\vh}}_{\bar{\va}}= \mv(\theta)\tilde{\vh}_{\bar{\va}}$ the parts of the channel along the dominant directions and in the subspace orthogonal to the dominant direction, respectively.
In~\cite{Goldstein1997} two fast algorithms to find $\mv(\theta)$ are described. If we choose the transformation $\mt^\He(\theta)$ as unitary, the statistics of the noise $\vn$ are not changed.
As stated in~\cite{Griffiths1982}, the \ac{MMSE} is preserved for any full-rank transformation $\mt^\He(\theta)$.
We can now formulate two independent estimators for each part via
\begin{align}
	&\hat{{\vh}}_\va = \va(\theta)\mathbb{E}[\tilde{h}_\va\mid \Tilde{y}_\va,\theta], \\
	&\hat{{\vh}}_{\bar{\va}} = \mv(\theta)\mathbb{E}[\tilde{\vh}_{\bar{\va}}\mid \Tilde{\vy}_{\bar{\va}},\theta],
	\label{eq:cme_nlos}
\end{align}
The overall estimate is given as
\begin{align}
	\hat{\vh} =  \hat{{\vh}}_\va + \hat{{\vh}}_{\bar{\va}}.
\end{align}

\subsection{Estimation of Dominant Direction Component}
\label{sec:LoS}

The first entry of $\Tilde{\vy}$ defined in~\eqref{eq:transform} is
\begin{align}
	\Tilde{y}_\va &= 
	\va^\He(\theta) \vh + \va^\He(\theta) \vn = \Tilde{h}_\va + \Tilde{n},
\end{align}
where $\Tilde{h}_\va$ denotes the power of the channel along the steering vector $\va(\theta)$.
Due to the absence of a closed-form optimal solution, we utilize a \ac{LMMSE} estimator to obtain an estimate of ${\vh}_\va$ via
\begin{align}
	\hat{{\vh}}_{\va,\text{LMMSE}} &= \va(\theta)\frac{\mathbb{E}\left[|\Tilde{h}_\va|^2\right]}{ \mathbb{E}\left[|\Tilde{h}_\va|^2\right] + \sigma^2} \Tilde{y}_\va = \va(\theta)\frac{P_\va}{ P_\va + \sigma^2} \Tilde{y}_\va.\label{eq:lmmse_scalar}
\end{align}

\subsection{Estimation of Orthogonal Subspace Component}

The \ac{CME} for the second component given in~\eqref{eq:cme_nlos} can not be calculated analytically, since the \acp{PDF} of the distributions are unknown.
To this end, let us first observe the following
\begin{align}
	f_{\Tilde{\vh}_{\bar{\va}}\mid \Tilde{\vy}_{\bar{\va}},\theta}(\Tilde{\vh}_{\bar{\va}}\mid \Tilde{\vy}_{\bar{\va}},\theta) &= \frac{f_{\Tilde{\vy}_{\bar{\va}}\mid \Tilde{\vh}_{\bar{\va}},\theta}(\Tilde{\vy}_{\bar{\va}}\mid \Tilde{\vh}_{\bar{\va}},\theta)f_{\Tilde{\vh}_{\bar{\va}}\mid \theta}(\Tilde{\vh}_{\bar{\va}}\mid\theta)}{f_{\Tilde{\vy}_{\bar{\va}}\mid \theta}(\Tilde{\vy}_{\bar{\va}}\mid\theta)} \\
	&= \frac{f_{\Tilde{\vn}}(\Tilde{\vy}_{\bar{\va}}- \Tilde{\vh}_{\bar{\va}})f_{\Tilde{\vh}_{\bar{\va}}\mid \theta}(\Tilde{\vh}_{\bar{\va}}\mid\theta)}{f_{\Tilde{\vy}_{\bar{\va}}\mid \theta}(\Tilde{\vy}_{\bar{\va}}\mid\theta)}.
\end{align}
If $f_{\Tilde{\vh}_{\bar{\va}}\mid \theta}(\Tilde{\vh}_{\bar{\va}}\mid\theta)$ and $f_{\Tilde{\vy}_{\bar{\va}}\mid \theta}(\Tilde{\vy}_{\bar{\va}}\mid\theta)$ are distributed following a \ac{GMM}, a closed form solution for the \ac{CME} $\mathbb{E}[\tilde{\vh}_{\bar{\va}}\mid \Tilde{\vy}_{\bar{\va}},\theta]$ can be formulated, cf.,~\cite{Koller2022}.

To this end, we approximate the \ac{PDF} of $\vh_{\bar{\va}}$ using a \ac{GMM} motivated by their universal approximation property~\cite{Koller2022, Nguyen2020}
\begin{align}
	f_{\vh_{\bar{\va}}}^{(J)}(\vh_{\bar{\va}}) = \sum_{j=1}^{J} p(j) \NC(\vh_{\bar{\va}};\vmu_j,\mc_j), \label{eq:GMM_nlos}
\end{align}
where $p(j)$, $\vmu_j$, and $\mc_j$ are the mixing coefficient, mean,
and covariance matrix of the $j$-th \ac{GMM} component, respectively.
Let us further introduce the following approximation
\begin{align}
	{\vh}_{\bar{\va}} &= (\I - \va(\theta)\va^\He(\theta))\vh\\
	&= \sqrt{\frac{M}{K+1}} \vh_\text{NLoS} - \sqrt{\frac{M}{K+1}}\va(\theta)\va(\theta)^\He \vh_\text{NLoS} \label{eq:proj_sum}\\
	&\approx \sqrt{\frac{M}{K+1}} \vh_\text{NLoS}, \label{eq:nlos_approx}
\end{align}
where we essentially neglect the contribution of the second term in~\eqref{eq:proj_sum}. 
This assumption is reasonable when the number of antennas is large enough~\cite{Bjoernson2017}.
Building on the approximation of $\vh_{\bar{\va}}$ in~\eqref{eq:nlos_approx} and assuming that it is uncorrelated with the \ac{DoA} $\theta$, i.e., $\vh_{\bar{\va}} \mid \theta = \vh_{\bar{\va}}$, we can now obtain an expression for the conditional \ac{PDF} of $\Tilde{\vh}_{\bar{\va}}$ based on~\eqref{eq:GMM_nlos} as
\begin{align}
	f^{(J)}_{\Tilde{\vh}_{\bar{\va}}\mid \theta}(\Tilde{\vh}_{\bar{\va}}\mid\theta) &= \sum_{j=1}^{J} p(j) \NC(\vh_{\bar{\va}};\mv(\theta)^\He\vmu_j,\mv(\theta)^\He\mc_j\mv(\theta)),\label{eq:GMM_cond_nlos}
\end{align}
which follows from the transformation $\mt^\He(\theta)$ in~\eqref{eq:transform}.
The conditional \ac{PDF} of the pilot observation $\Tilde{\vy}_{\bar{\va}}$ can then be approximated by
\begin{multline}
	f_{\Tilde{\vy}_{\bar{\va}}\mid\theta}^{(J)}(\Tilde{\vy}_{\bar{\va}}\mid\theta) = \\\sum_{j=1}^{J} p(j) \NC(\Tilde{\vy}_{\bar{\va}};\mv(\theta)^\He\vmu_j,\mv(\theta)^\He\mc_j\mv(\theta) + \mc_{\Tilde{\vn}}).
	\label{eq:GMM_cond_noisey_nlos}
\end{multline}

Leveraging this found approximation of the conditional \acp{PDF} of $\Tilde{\vh}_{\bar{\va}}$ and $\Tilde{\vy}_{\bar{\va}}$, a convex combination of LMMSE estimates can be used to find the channel estimate in~\eqref{eq:cme_nlos}~\cite{Koller2022}.
As the covariance matrices of each component in~\eqref{eq:GMM_cond_nlos} and~\eqref{eq:GMM_cond_noisey_nlos} are calculated using the current \ac{DoA}, we propose to utilize an alternative approach, which was introduced in~\cite{Weisser2023a}, to reduce computational complexity.
Here, instead of solving the estimation for $\Tilde{\vh}_{\bar{\va}}$ we directly estimate ${\vh}_{\bar{\va}}$.
This is done by taking the projection $\vy_{\bar{\va}} = \mv(\theta)\mv^\He(\theta)\vy$ as a preprocessing step and consequently leaving the $J$ \ac{LMMSE} filters unaltered.
Following a similar argumentation as in \cite{Weisser2023a}, we assume that the noise of the $\vy_{\bar{\va}}$ component is Gaussian with the covariance matrix $\mc_{\bar{\vn}} = \sigma^2\frac{M-1}{M}\I_M$.
The resulting estimator can be written as
\begin{align}
	\hat{\vh}_{\bar{\va},\text{GMM}} = \sum_{j=1}^J p(j\mid{\vy}_{\bar{\va}})
	\hat{{\vh}}_{\bar{\va},\text{GMM},j}\label{eq:proj1}
\end{align}
with
\begin{align}
	p(j\mid{\vy}_{\bar{\va}}) = \frac{p(j)\NC\left({\vy}_{\bar{\va}};\vmu_j,\mc_j + \mc_{\bar{\vn}}\right)}{\sum_{i=1}^J p(i)\NC\left({\vy}_{\bar{\va}};\vmu_i,\mc_i + \mc_{\bar{\vn}}\right)}\label{eq:proj2}
\end{align}
and
\begin{align}
	\hat{{\vh}}_{\bar{\va},\text{GMM},j} = \mc_j\left(\mc_j + \mc_{\bar{\vn}}\right)^{-1}\left({\vy}_{\bar{\va}} - \vmu_j\right) + \vmu_j.\label{eq:proj3}
\end{align}
As our simulations did not show any significant difference between the estimator based on~\eqref{eq:GMM_cond_nlos} and~\eqref{eq:GMM_cond_noisey_nlos} and the computational more efficient estimator formulated in~(\ref{eq:proj1})-(\ref{eq:proj3}), we will focus in the following on the latter, due to better complexity.

\subsection{Augmentation of Training Data}

In order to approximate the proposed estimators based on a representative training set $\mathcal{H}=\{\vh_{\ell}\}_{\ell=1}^L$ of $L$ samples, 
the following augmentation is done.
First, we determine the \ac{DoA}~$\theta_\ell$ of each channel realization $\vh_\ell$ within the training set $\mathcal{H}$ as
\begin{align}
	\theta_\ell = \arg \max_{\theta \in \Theta} |\va^\He(\theta)\vh_\ell|, \label{eq:doa_est}
\end{align}
where $\Theta$ defines the set of possible angles.
Sequentially,
we construct two sets $\mathcal{H}_{\va}=\{\tilde{h}_{{{\va}},\ell}\}_{\ell=1}^L$ and $\mathcal{H}_{\bar{\va}}=\{\vh_{{\bar{\va}},\ell}\}_{\ell=1}^L$,
where we project each channel sample $\vh_\ell$ using the transformation $\mt(\theta_\ell)$.
This is done with
\begin{align}
	\tilde{h}_{{{\va}},\ell} = \va^\He(\theta_\ell)\vh_{\ell}
\end{align}
and
\begin{align}
	\vh_{{\bar{\va}},\ell} = \mv(\theta_\ell)\mv^\He(\theta_\ell)\vh_{\ell} = \left(\I_M - \va(\theta_\ell)\va^\He(\theta_\ell)\right)\vh_{\ell},
\end{align}
where $\theta_\ell$ is provided by \eqref{eq:doa_est}.
Based on the training set $\mathcal{H}_\va$ we  can estimate $\hat{P}_\va = 1/L\sum_{\ell=1}^L |\tilde{h}_{\va,\ell}|^2$ in~\eqref{eq:lmmse_scalar} and given $\mathcal{H}_{\bar{\va}}$, the fitting of the components in \eqref{eq:GMM_nlos} is done with the well-known expectation-maximization (EM) algorithm~\cite{Bishop2006}.

\subsection{DoA Estimation}

The transformation $\mt(\theta)$ in~\eqref{eq:transform} is based on the \ac{DoA} of the \ac{LOS} path. 
As in practice, the true angle is not given, we approximate the transform with
\begin{align}
	\mt(\theta) \approx \mt(\hat{\theta}),
\end{align}
where $\hat{\theta}$ denotes the \ac{DoA} estimate.
In a \ac{JCAS} setup, the \ac{DoA} estimate or the user's position might be given by other system functionalities, e.g., radar. 
Since this assumption may not always be fulfilled, especially regarding systems not directly built for \ac{JCAS}, the \ac{DoA} estimation may need to be conducted while estimating the channel. 
We want to emphasize, that our proposed approach is independent of the used \ac{DoA} estimation algorithm, e.g., the well-known \ac{MUSIC} algorithm~\cite{Schmidt1986}, root-\ac{MUSIC}~\cite{Barabell1983} and \ac{ESPRIT}~\cite{Roy1989}.
Additionally, as the \acp{DoA} do not change rapidly, computational efficient tracking algorithms, which either track the subspace~\cite{Yang1995, Utschick2002} or the \ac{DoA}~\cite{Sastry1991}, can be utilized.

\section{Baseline estimators}

The conventional \ac{LS} channel estimator is given as $\hat{\vh}_\text{LS} = \vy$.
Furthermore, the \ac{GMM} estimator from~\cite{Koller2022} can be used to estimate the whole channel by fitting the GMM onto the representative data set $\mathcal{H}=\{\vh_\ell\}_{\ell=1}^L$ and inferring the estimate $\hat{\vh}_\text{GMM}$.

The \ac{OMP}~\cite{Pati1993} is a well-known example from the field of \ac{CS}.
Since it needs to know the sparsity order, we evaluate a genie-aided variant (``genieOMP''), where \ac{OMP} chooses the optimal sparsity order based on the true channel $\vh$.

Purely parametric channel estimation schemes, e.g., \cite{Shafin2016}, assume a ray-based channel model
as $\vh\approx \ma(\vth)\vs$, where $\ma(\vth)=[\va(\theta_1),...,\va(\theta_P)] \in \mathbb{C}^{M\times P}$ is the steering matrix of the $P$ paths and $\vs$ holds the complex path gains.
For a given set of estimated \acp{DoA} $\hat{\vth}$, the \ac{LS} solution can be found as
\begin{align}
	\hat{\vh}_\text{para} = \ma(\hat{\vth})\left(\ma^\He(\hat{\vth})\ma(\hat{\vth})\right)^{-1}\ma^\He(\hat{\vth}) \vy.
\end{align}
Similar to~\cite{Shafin2016}, we implement forward-backward averaging for the \ac{ESPRIT} algorithm to enhance its estimation quality.
As the number of paths needs to be known, we again implement genie-aided variants, 
where the number of paths is chosen based on the true channel $\vh$. 

Another estimator
is based on the global sample covariance matrix, which we can compute from the training data set $\mathcal{H}_{\bar{\va}}$ with 
$
	\mc_{\bar{\va}} = 
 1/L
 \sum_{\vh_{\bar{\va}}\in \mathcal{H}_{\bar{\va}}} \vh_{\bar{\va}}\vh_{\bar{\va}}^\He.
$
We can use this sample covariance matrix to formulate the \ac{LMMSE} estimator for the \ac{NLOS} components as
\begin{align}
	\hat{\vh}_{\bar{\va},\text{s-cov}} =\mc_{\bar{\va}} \left(\mc_{\bar{\va}} + \sigma^2\frac{M-1}{M}\I_M\right)^{-1}\matp_{\bar{\va}}\vy.
\end{align}
The \ac{LOS} component is estimated as in \eqref{eq:lmmse_scalar}.

When working with the channel model described in Section \ref{sec:system_model}, the true conditional covariance matrix $\mc_{\vdel,\theta}$ of every channel realization is known, and we can construct the utopian genie \ac{LMMSE} estimator as
\begin{align}
	\hat{\vh}_\text{genieLMMSE} &= \mc_{\vdel,\theta}\left(\mc_{\vdel,\theta} + \sigma^2\I_M\right)^{-1} \vy.
\end{align}

\section{Numerical Simulations}

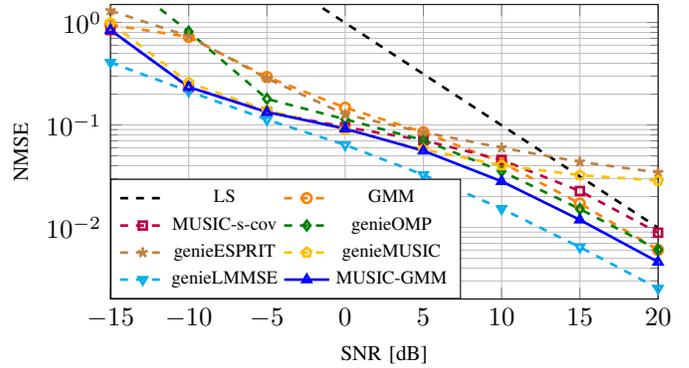
\begin{figure}
		\centering
		\begin{tikzpicture}
			\centering
			\begin{semilogyaxis}[
				width=0.985\columnwidth,
				height=5.5cm,
				ylabel={\footnotesize NMSE},
				xlabel={\footnotesize SNR [dB]},
				xmin=-15,
				xmax=20,
				ymin=0.002,
				ymax=1.5,
				grid=both,
				xtick={-20,-15,-10,-5,0,5,10,15,20},
				ytickten={-6,...,3},
				legend columns=2,
				legend style={at={(0,0)}, anchor=south west, font=\scriptsize},
				]		
				
				\addplot[mark size=2pt, dashed,line width=1pt, color=black,mark options={solid}] table [col sep=comma] {data/nmse_32_3gpp_LS_10snaps_3clusters_rician10.csv};
				\addlegendentry{LS}
				\addplot[mark=o,mark size=1.8pt, dashed,line width=1pt, color=orange,mark options={solid}] table [col sep=comma] {data/nmse_64_gmm_3gpp_gmm_10snaps_3clusters_rician10.csv};
				\addlegendentry{GMM}
				\addplot[mark=square,mark size=1.5pt, dashed,line width=1pt, color=purple,mark options={solid}] table [col sep=comma] {data/nmse_64_ge_3gpp_DoA_smpl_cov_10snaps_3clusters_rician10.csv};
				\addlegendentry{MUSIC-s-cov}
				\addplot[mark=diamond,mark size=1.8pt, dashed,line width=1pt, color=green!50!black,mark options={solid}] table [col sep=comma] {data/nmse_64_3gpp_OMP_10snaps_3clusters_rician10.csv};
				\addlegendentry{genieOMP}
				\addplot[mark=star,mark size=2pt, line width=1pt,dashed, color=brown,mark options={solid}] table [col sep=comma] {data/nmse_64_esprit_3gpp_genieESPRIT_10snaps_3clusters_rician10.csv};
				\addlegendentry{genieESPRIT}
				\addplot[mark=pentagon,mark size=1.8pt, line width=1pt,dashed, color=orange!50!yellow,mark options={solid}] table [col sep=comma] {data/nmse_64_music_3gpp_genieMUSIC_10snaps_3clusters_rician10.csv};
				\addlegendentry{genieMUSIC}
				\addplot[mark=triangle,mark size=1.8pt, line width=1pt,dashed, color=cyan,mark options={solid, rotate=180}] table [col sep=comma] {data/nmse_64_ge_3gpp_genie_10snaps_3clusters_rician10.csv};
				\addlegendentry{genieLMMSE}
				
				\addplot[mark=triangle,mark size=1.8pt, line width=1pt, color=blue,mark options={solid}] table [col sep=comma] {data/nmse_64_ge_3gpp_DoA_est_10snaps_3clusters_rician10.csv};
				\addlegendentry{MUSIC-GMM}

			\end{semilogyaxis}
		\end{tikzpicture}

   \vspace{-2mm}
	\caption{NMSE for given channel estimations based on $T=10$ pilot observations. The angular spread of the \ac{NLOS} clusters set to $\sigma_\text{AS}=2^\circ$ and the Rician factor is $K=10$ dB.}
  \vspace{-2mm}
	\label{fig:nmse_3gpp_SNR_3clus}
\end{figure}

In this work, we consider a base station that serves a sector of $120^\circ$, i.e., the angle of the \ac{LOS} path is drawn uniformly in the interval $[-60^\circ,60^\circ]$.
We normalize the channel realizations with $\mathbb{E}\left[\|\vh\|^2\right]=M$ such that we can define the \ac{SNR} $=\frac{1}{\sigma^2}$.
Given $N$ channel estimates $\{\hat{\vh}_n\}^N_{n=1}$ of the test samples $\{\vh_n\}^N_{n=1}$, we can define the \ac{NMSE} as $\frac{1}{NM}\sum_{n=1}^N \|\vh_n - \hat{\vh}_n\|^2$.
The simulations are carried out $N = 10^3$ times. 
We use $T=10$ observation for each covariance coherence interval if not stated otherwise.
The number of \ac{BS} antennas is $M=64$ with an antenna spacing of $d=\lambda/2$.
We simulate three \ac{NLOS} propagation clusters, assuming an angular spread of $\sigma_\text{AS}=2^\circ$ in accordance to~\cite{3GPP2020}, if not stated otherwise.
The \ac{NLOS} channel realizations are then constructed according to~\cite{Neumann2018}.
We use $1.5\cdot10^5$ training samples to fit a \ac{GMM} with $J=128$ components for each scenario, where we use a grid $\Theta$ with $16M$ points to estimate the \ac{DoA}.
For the first stage of the proposed method, we use \ac{MUSIC}, with the same grid size.
To this end, we take the last $T$ observations and construct the sample covariance matrix
\begin{align}
	\hat{\mc}_\vy = \frac{1}{T} \my\my^\He, \label{eq:spaceCov}
\end{align}
where $\my=[\vy(1),...,\vy(T)] \in \mathbb{C}^{M\times T}$.
Based on $\hat{\mc}_\vy$, we calculate the \ac{DoA} estimate $\hat{\theta}$.
The resulting estimator is denoted as ``MUSIC-GMM''.
The ``MUSIC-s-cov'' estimator utilizes the same training samples.

In Fig.~\ref{fig:nmse_3gpp_SNR_3clus}, the performance for a Rician factor of $K=10$~dB is shown for different \ac{SNR}.
We can observe that in the low \ac{SNR} regime, the three methods based on \ac{MUSIC} perform the best, even approaching the ``genieLMMSE''.
For this low \ac{SNR} region, the estimation of the \ac{LOS} component dominates the problem and, hence, all \ac{MUSIC}-based estimators perform similar.
For high \ac{SNR} values, estimating the \ac{NLOS} clusters is important. 
Here, our proposed ``MUSIC-GMM'' estimator
performs extraordinarily well.
Furthermore, the parametric channel estimators saturate for high \ac{SNR} due to their model mismatch.

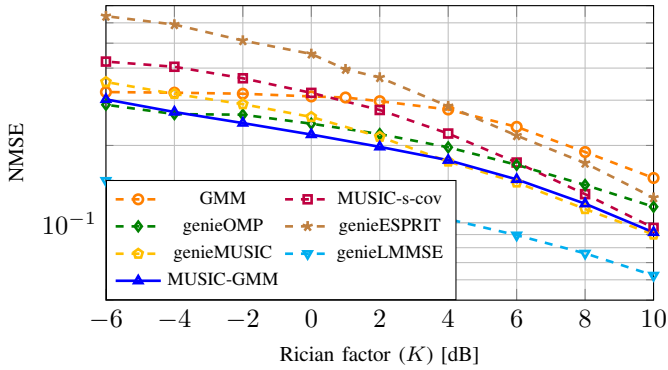
\begin{figure}
	\centering
	\begin{tikzpicture}
		\centering
		\begin{semilogyaxis}[
			width=0.985\columnwidth,
			height=5.5cm,
			ylabel={\footnotesize NMSE},
			xlabel={\footnotesize Rician factor ($K$) [dB]},
			xmin=-6,
			xmax=10,
			ymin=0.05,
			ymax=0.7,
			grid=both,
			xtick={-10,-8,-6,-4,-2,0,2,4,6,8,10},
			ytickten={-6,...,3},
			legend columns=2,
			legend style={at={(0,0)}, anchor=south west, font=\scriptsize},
			]		
			
			\addplot[mark=o,mark size=1.8pt, dashed,line width=1pt, color=orange,mark options={solid}] table [col sep=comma] {data/nmse_64_gmm_3gpp_gmm_10snaps_0dB_3clusters_rician.csv};
			\addlegendentry{GMM}
			\addplot[mark=square,mark size=1.5pt, dashed,line width=1pt, color=purple,mark options={solid}] table [col sep=comma] {data/nmse_64_ge_3gpp_DoA_smpl_cov_10snaps_0dB_3clusters_rician.csv};
			\addlegendentry{MUSIC-s-cov}
			\addplot[mark=diamond,mark size=1.8pt, dashed,line width=1pt, color=green!50!black,mark options={solid}] table [col sep=comma] {data/nmse_64_3gpp_OMP_10snaps_0dB_3clusters_rician.csv};
			\addlegendentry{genieOMP}
			\addplot[mark=star,mark size=2pt, line width=1pt,dashed, color=brown,mark options={solid}] table [col sep=comma] {data/nmse_64_esprit_3gpp_genieESPRIT_10snaps_0dB_3clusters_rician.csv};
			\addlegendentry{genieESPRIT}		
			\addplot[mark=pentagon,mark size=1.8pt, line width=1pt,dashed, color=orange!50!yellow,mark options={solid}] table [col sep=comma] {data/nmse_64_music_3gpp_genieMUSIC_10snaps_0dB_3clusters_rician.csv};
			\addlegendentry{genieMUSIC}
			\addplot[mark=triangle,mark size=1.8pt, line width=1pt,dashed, color=cyan,mark options={solid, rotate=180}] table [col sep=comma] {data/nmse_64_ge_3gpp_genie_10snaps_0dB_3clusters_rician.csv};
			\addlegendentry{genieLMMSE}
			\addplot[mark=triangle,mark size=1.8pt, line width=1pt, color=blue,mark options={solid}] table [col sep=comma] {data/nmse_64_ge_3gpp_DoA_est_10snaps_0dB_3clusters_rician.csv};
			\addlegendentry{MUSIC-GMM}

		\end{semilogyaxis}
	\end{tikzpicture}
 
 \vspace{-2mm}
	\caption{
		NMSE for given channel estimations based on $T=10$ pilot observations at SNR $=0$ dB. The angular spread of the \ac{NLOS} clusters is set to $\sigma_\text{AS}=2^\circ$.	
}
\vspace{-2mm}
	\label{fig:nmse_3gpp_rician}
\end{figure}

Since the channel characteristics depend heavily on the Rician factor $K$, Fig.~\ref{fig:nmse_3gpp_rician} shows the performance for different Rician factors.
All considered estimators show increasing performance with an increase of the Rician factor. 
This is because estimating the \ac{NLOS} components becomes less significant.
Although ``genieOMP'' and ``genieMUSIC'' utilize genie knowledge, our proposed channel estimation method exhibits at least a comparable performance for the different Rician factors.

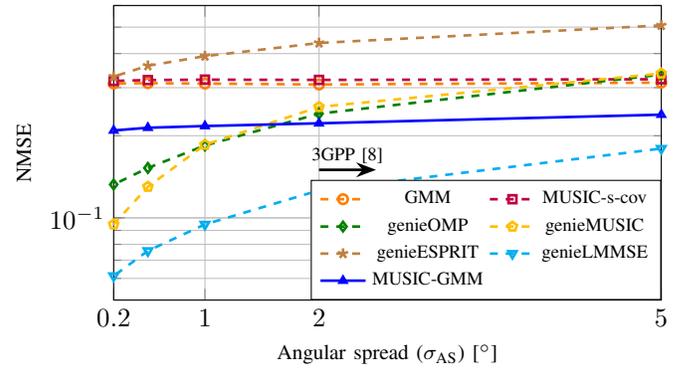
\begin{figure}
	\centering
	\begin{tikzpicture}
		\centering
		\begin{semilogyaxis}[
			width=0.985\columnwidth,
			height=5.5cm,
			ylabel={\footnotesize NMSE},
			xlabel={\footnotesize Angular spread ($\sigma_\text{AS}$) [$^\circ$]},
			xmin=0.2,
			xmax=5,
			ymin=0.05,
			ymax=0.6,
			grid=both,
			xtick={0.01,0.02,0.05,0.1, 0.2, 1,2,5,10},
			xticklabels={$0.01$,$0.02$,$0.05$,$0.1$,$0.2$, $1$,$2$,$5$,$10$},
			ytickten={-6,...,3},
			legend columns=2,
			legend style={at={(1,0)}, anchor=south east, font=\scriptsize},
			]		

			\addplot[mark=o,mark size=1.8pt, dashed,line width=1pt, color=orange,mark options={solid}] table [col sep=comma] {data/nmse_64_gmm_3gpp_gmm_10snaps_0dB_3clusters_rician0_spread.csv};
			\addlegendentry{GMM}
			\addplot[mark=square,mark size=1.5pt, dashed,line width=1pt, color=purple,mark options={solid}] table [col sep=comma] {data/nmse_64_ge_3gpp_DoA_smpl_cov_10snaps_0dB_3clusters_rician0_spread.csv};
			\addlegendentry{MUSIC-s-cov}
			\addplot[mark=diamond,mark size=1.8pt, dashed,line width=1pt, color=green!50!black,mark options={solid}] table [col sep=comma] {data/nmse_64_3gpp_OMP_10snaps_3clusters_rician0_spread.csv};
			\addlegendentry{genieOMP}
			\addplot[mark=pentagon,mark size=1.8pt, line width=1pt,dashed, color=orange!50!yellow,mark options={solid}] table [col sep=comma] {data/nmse_64_music_3gpp_genieMUSIC_10snaps_0dB_3clusters_rician0_spread.csv};
			\addlegendentry{genieMUSIC}
			\addplot[mark=star,mark size=2pt, line width=1pt,dashed, color=brown,mark options={solid}] table [col sep=comma] {data/nmse_64_esprit_3gpp_genieESPRIT_10snaps_0dB_3clusters_rician0_spread.csv};
			\addlegendentry{genieESPRIT}
			\addplot[mark=triangle,mark size=1.8pt, line width=1pt,dashed, color=cyan,mark options={solid, rotate=180}] table [col sep=comma] {data/nmse_64_ge_3gpp_genie_10snaps_0dB_3clusters_rician0_spread.csv};
			\addlegendentry{genieLMMSE}
			\addplot[mark=triangle,mark size=1.5pt, line width=1pt, color=blue,mark options={solid}] table [col sep=comma] {data/nmse_64_ge_3gpp_DoA_est_10snaps_0dB_3clusters_rician0_spread.csv};
			\addlegendentry{MUSIC-GMM}
			
		\draw [-{Stealth}, line width=1pt] (2,0.15)--(2.5,0.15);
        \node at (2.25,0.17) {\scriptsize 3GPP~\cite{3GPP2020}};
		\end{semilogyaxis}
	\end{tikzpicture}
 
 \vspace{-2mm}
	\caption{
		NMSE for given channel estimations based on $T=10$ pilot observations at SNR $=0$ dB and $K=0$ dB.		
}
\vspace{-2mm}
	\label{fig:nmse_3gpp_spread}
\end{figure}

Fig.~\ref{fig:nmse_3gpp_spread} shows the dependency of the performance on the angular spread $\sigma_\text{AS}$  of the \ac{NLOS} clusters. 
As $\sigma_\text{AS}$ increases, the parametric and sparse channel estimators degrade in their performance. 
The angular spread for a typical sub $6$~GHz scenario is specified in~\cite{3GPP2020} as $\sigma_\text{AS}=2^\circ$ and $\sigma_\text{AS}=5^\circ$ for urban macro and urban micro cells, respectively.
With increasing angular spread $\sigma_\text{AS}$, our proposed method exhibits a smaller gap to the utopian ``genieLMMSE'' and outperforms all of the other baselines for $\sigma_\text{AS}\geq2^\circ$.

\section{Conclusion}

In this work, we proposed a \ac{DoA}-aided \ac{MMSE}-based channel estimation.
To this end, we first introduced a \ac{DoA}-based reformulation of the observations similar to \ac{GSC}.
Two independent estimators are introduced, where the \ac{NLOS} clusters are estimated utilizing the \ac{GMM} framework from~\cite{Koller2022}. 
Extensive simulations showed superior performance of our proposed estimation approach compared to state-of-the-art baselines in the considered realistic scenarios. 
Further evaluations with real-world measurements are planned for future work.
Additionally, we intend to investigate further applications, including coarsely quantized systems and the movement of \acp{UE} along trajectories.

\bibliographystyle{IEEEtran}
  
\bibliography{IEEEabrv,mybib}

\end{document}